\documentclass[aps,prev,twocolumn,preprintnumbers,floatfix,nofootinbib]{revtex4-1}
\pdfoutput=1
\usepackage{graphicx}
\usepackage{bm}
\usepackage{times}
\usepackage{slashed}
\usepackage{color}
\usepackage{color}
\usepackage{amsmath}

\begin{document}

%\title{A 1 GeV WIMP is Not a Good Way to Explain the DAMA Result}
\title{GeV WIMPs scattering off of OH impurities cannot explain the DAMA signal}

\author{Stefano Profumo}\email{profumo@ucsc.edu} 
\author{Farinaldo S. Queiroz}\email{fdasilva@ucsc.edu}

%\vspace{1cm}
\affiliation{Department of Physics and Santa Cruz Institute for Particle Physics
University of California, Santa Cruz, CA 95064, USA
}

\date{\today}
\vspace{1cm}

\begin{abstract}

In the presence of OH impurities in the DAMA crystals, GeV-scale WIMPs elastically scattering off of hydrogen nuclei with a spin independent cross section of $\sim 10^{-33}\ {\rm cm}^2$ might explain the annual modulation observed by the DAMA experiment, while being consistent with other direct dark matter searches, as scattering would occur at energies below the energy threshold of other detectors. In this work we examine this possibility and show that, independent of the level of OH impurities in the DAMA crystals, for several reasons this scenario  does not provide a viable explanation to the DAMA signal. %Since, up to date, no background has been identified with a phase, spectrum and rate compatible with DAMA/LIBRA’s signal, this modulation remains a mystery that requires a more exotic explanation. 
\end{abstract}

\maketitle

\section{Introduction}
The particle nature of the dark matter (DM) remains one of the most troubling mysteries of our universe. Weakly interacting massive particles, or WIMPs, are for many reasons leading particle dark matter candidates, and, as such, they are being very actively searched in a broad variety of experiments. In the recent past, a variety of puzzling signals have been gathered from several direct detection experiments, tentatively pointing, if interpreted as originating from WIMP-nuclei scattering, towards a mass window in the 5-10~GeV range. Such WIMPs, light compared to standard theory-driven expectations, might for example provide an interpretation to the excess events reported by CDMS-SI \cite{Agnese:2013rvf}, CoGeNT \cite{Aalseth:2010vx,Aalseth:2011wp,cogentnew}, CRESST \cite{Angloher:2011uu} and DAMA \cite{Bernabei:2013qx}. 

Whether or not all the putative  signals listed above can be made consistent with each other is matter of some contention \cite{weiner,cogentnew}. More importantly, null results from the XENON \cite{xenon} and LUX \cite{LUX} collaborations are entirely incompatible, at last at face value, with a light WIMP interpretation, with the possible exception of highly tuned iso-spin violation scenarios \cite{isospin}, or if the scintillation properties of LXe are fiercely changed from the assumptions used by the LUX experiment \cite{LUX}. Additionally, the most recent CDMSlite \cite{cdmslite} and SUPERCDMS \cite{supercdms} that use the same target nucleus strongly strongly disfavor these light dark matter ($7-10$~GeV) positive signals.

Recently, it was pointed out in Ref.~\cite{claim}  that a $\mathcal{O}(1)$~GeV WIMP might provide a possible explanation to the DAMA signal consistent with other direct detection experiments. The key point made in Ref.~\cite{claim} is that a contamination of OH-molecules is likely to be present in the NaI(Tl) crystals used by the DAMA collaboration. The precise level of such contamination is unknown, but, if at all at a reasonable level, it would be supposedly sufficient to allow a $\mathcal{O}(1)$~GeV WIMP to strike a Hydrogen nucleus only, which in turn would produce photons then captured by the PMTs in the DAMA detector. Ref.~\cite{claim} hypothesizes and utilizes a $1$~ppm contamination level; In what follows we will assume this same value. 

In this study we explore whether OH impurities indeed lead to a self-consistent explanation of the DAMA signal, focussing on the broad very light WIMP (for definiteness, $1-8$~GeV) window, and thus extending the mass range originally under consideration in Ref.~\cite{claim}. We also study the effect of a variation in the assumptions about the WIMP velocity distribution in the Galaxy. We conclude that in general OH impurities do not provide a viable scenario to explain the DAMA signal, due to a variety of constraints, including residual scattering off of Na nuclei, results from high-altitude detectors, and unacceptably large contributions to the heat flow of the Earth.
% Further, we will briefly show that this proposal seems to be, at least, unlikely.
%

\section{Preliminaries}
%
%With the present technology, the annual modulation is the main model independent signature for the dark matter signal. Despite the fact that is expected to be relatively small, a suitable large-mass, low-radioactive set-up with an efficient control of the running conditions can point out its presence. 
Direct detection of dark matter refers to measurements of the energy recoil of a WIMP-Nucleus scattering event. Because of the Earth orbital motion an annual modulation in the WIMPs scatterings rate in underground detectors generically exists. In the elastic dark matter setup we can define the differential scattering rate per unit detector mass as:
\begin{equation}
\label{dRdE}
  \frac{dR}{dE_{nr}} = \frac{2\rho_{\chi}}{m_{\chi}}
      \int d^3v \, v f(\bf{v},t) \frac{d\sigma}{dq^2}(q^2,v) \, ,
\end{equation}where $q$ is the momentum transfer, $\rho_{\chi}\simeq0.3\ {\rm GeV/cm^3}$ is the local dark matter density, $f(\bf{v},t)$ is the local dark matter velocity distribution, which we assume to be time-independent and of the functional form
\begin{equation}
 \label{veloSHM}
f_{SHM}(u) = \begin{cases} \frac{1}{N v_0^3\pi^{3/2}} e^{-u^2/v_0^2} &\mbox{if } u < v_{esc} \\ 
0 & \mbox{otherwise}, \end{cases} 
\end{equation}and, finally, where $\frac{d\sigma}{dq^2}(q^2,v)$ is the differential WIMP-Nucleus scattering cross section given by,
\begin{equation}\label{eqn:dsigmadq}
  \frac{d\sigma}{dq^2}(q^2,v)
    = \frac{\sigma_{SI}}{4 \mu^2 v^2} F^2(q),
\end{equation}with $F^2(q)$ the Helm form factor which accounts for the finite size of the nucleus, $\mu$  the WIMP-nucleus reduced mass, and $\sigma_{SI}$  the spin independent WIMP-nucleon scattering cross section. The latter can be cast as,
\begin{equation} \label{eqn:sigmaSI2}
  \sigma_{SI} = \frac{4}{\pi} \mu^2
             \left[Z f_p + (A-Z) f_n \right]^{2},
\end{equation}$f_{p},f_{n}$ being the effective couplings to proton and neutrons, respectively, $A$  the atomic mass number and $Z$ the atomic number. These effective couplings and the Helm form factor are determined following a well known procedure, described e.g. in \cite{modulationreview}. Since we are interested in the very light WIMP and energy region, the form factor dependence is modest \cite{modulationreview}, therefore different form factors induce very mild changes in our conclusions.

In this study we adopt a Standard Halo Model for the velocity distribution, given in Eq.(\ref{veloSHM}). We also discuss departures from this model, especially when it comes to changing the one crucial input quantity for low-mass WIMPs, i.e. the dark matter Galactic escape velocity $V_{\rm esc}$.

\section{Overview of the DAMA/LIBRA Experiment}
The DAMA/LIBRA experiment, located in the underground INFN Gran Sasso National Laboratory in Italy, makes use of a large mass detector of about 250 kg highly radio pure NaI(Tl). The experiment is designed to detect nuclear recoil events through scintillation light, and is optimized to search for time variations in the event rate, rather than to identify dark matter scattering on an event-by-event basis. The strategy of looking for an annual modulation can be used to distinguish a dark matter signal from most possible backgrounds sources, although one may worry about possible sources of background which could potentially also exhibit seasonal variation. 

The DAMA/LIBRA collaboration claims, however, that there exists no background satisfying at once all of the following criteria: 

(i) having a co-sinusoidal modulation rate; 

(ii) existing only in a definite low energy range;

(iii) having exactly a one year period; 

(iv) possessing the proper phase; 

(v) producing single-hit events; 

(vi) having the proper modulation amplitude. 

The DAMA/LIBRA collaboration has so far released results corresponding to a total exposure of 1.17 ton yr over 13 annual cycles supporting the detection of a signal possibly due to dark matter at the $\sim 9 \sigma$ level \cite{Bernabei:2013qx}.

Regarding detector details, it is important to point out that the collaboration uses gamma ray sources, which induce electron recoil, to calibrate their detectors. Therefore for a given electron recoil with energy $E_{ee}$ the DAMA experiment is calibrated to the resulting amount of light produced. When a nuclear recoil happens and some amount of energy is produced, DAMA thus quotes the electron equivalent energy that would have yielded the same amount of light. Based on this information, we can define a quenching factor ($Q$) which translates the ratio of the amount of scintillation light created by a nuclear recoil to that produced by a electron recoil of the same energy. The quenching factor value varies according to the target, and in our work, we will use the values reported by the collaboration, $Q_{Na}=0.3$ and $Q_{I}=0.09$. Moreover, we have accounted for the energy resolution of the detector by writing the differential rate in terms of the electron recoil energy as follows:
\begin{equation}
\label{eqn:dRdEee}
  \frac{dR}{dE_{ee}}(E_{ee},t)
    = \int_0^{\infty} dE_{nr} \,
       \phi(E_{nr},E_{ee}) \, \frac{dR}{dE_{nr}}(E_{nr},t),
\end{equation}where $\phi(E_{nr},E_{ee})$ is the differential response function and
$\phi(E_{nr},E_{ee})\,\Delta E_{ee}$ is the probability that a nuclear recoil
of energy $E_{nr}$ will produce a scintillation signal measured between
$E_{ee}$ and $E_{ee}+\Delta E_{ee}$, found to be
\begin{equation}\label{eqn:phi}
  \phi(E_{nr},E_{ee}) = \frac{1}{\sqrt{2\pi\sigma^2(QE_{nr})}}
                    e^{-(E_{ee}-Q E_{nr})^2/2\sigma^2(QE_{nr})},
\end{equation}with
\begin{equation}\label{eqn:sigmaER}
  \sigma(QE_{nr}) = \alpha\sqrt{QE_{nr}}+\beta\,QE_{nr},
\end{equation}
where $\alpha=(0.448\pm0.035)\sqrt{\rm keVee}$ and
$\beta=(9.1\pm5.1)\times 10^{-3}$.

As discussed in the previous section, due to the changing WIMP
velocity distribution at Earth as the Earth orbits the Sun, there is a
small ($\sim$ 1-10\%) variation (or modulation) in the recoil rate
throughout the year \cite{Bernabei:2013qx}.

The time
dependence in the recoil rate can be described as
\begin{equation} \label{eqn:dRdES}
  \frac{dR}{dE_{ee}}(E_{ee},t) = S_0(E_{ee}) + S_m(E_{ee}) \cos{\omega(t-t_0)} + \ldots
\end{equation}
where $S_0$ is the average rate, $S_m$ is the modulation amplitude,
and higher-order terms are neglected due to the small modulation amplitude, in the
absence of streams \cite{streamsDAMA}. We have packed the 36 bins reported by DAMA into 8 bins, in order to improve the sensitivity of statistical tests, according to Table I below.
\begin{table}[b]
  \addtolength{\tabcolsep}{1em}
  \begin{tabular}{c@{\hspace{1em}}c}
    \hline\hline
    Energy  & Average $S_m$ \\\empty
    [keVee] & [cpd/kg/keVee]  \\
    \hline
    2.0 -  2.5 & 0.0161 $\pm$ 0.0040 \\
    2.5 -  3.0 & 0.0260 $\pm$ 0.0044 \\
    3.0 -  3.5 & 0.0220 $\pm$ 0.0044 \\
    3.5 -  4.0 & 0.0084 $\pm$ 0.0041 \\
    4.0 -  5.0 & 0.0080 $\pm$ 0.0024 \\
    5.0 -  6.0 & 0.0065 $\pm$ 0.0022 \\
    6.0 -  7.0 & 0.0002 $\pm$ 0.0021 \\
    7.0 - 20.0 & 0.0005 $\pm$ 0.0006 \\
    \hline\hline
  \end{tabular}
  \caption{Average modulation amplitudes for the 8 bins used in this
  analysis. The original 36 bins of width 0.5~KeVee were packed into 8 as shown in the table.}
  \label{table1}
\end{table}

Using this data set, we have performed a chi-squared analysis, with the $\chi^2$ defined as follows:
\begin{equation}\label{eqn:chisquare}
  \chi^2(m_{\chi},\sigma)
    = \sum_i \frac{\left(S^{exp}_{m,i} - S_{m,i}^{th}(m_{\chi},\sigma)\right)^2}{\sigma_i^2} \, ,
\end{equation} where $S_{m,i}^{th} (m_{\chi},\sigma)$ is the theoretically expected amplitude for a WIMP model of a given mass ($m_{\chi}$) and spin independent scattering cross section ($\sigma$), whereas  $S_{m,i}^{\exp}$ is the experimental value of the modulation amplitude for a given bin i, and $\sigma_i$ is the uncertainty as indicated in Table~\ref{table1}.
%
%Now we have described how we the data analysis was done it is a good timing to present our results and investigate the very light WIMP window. Ref.~\cite{claim} states that the DAMA signal can be explained by a 1 GeV WIMP dark matter particle. We give here several arguments as to why this is not a viable possibility.
%

%
\begin{figure*}[t]
\centering
\mbox{\hspace*{-1cm}\includegraphics[width=0.73\columnwidth]{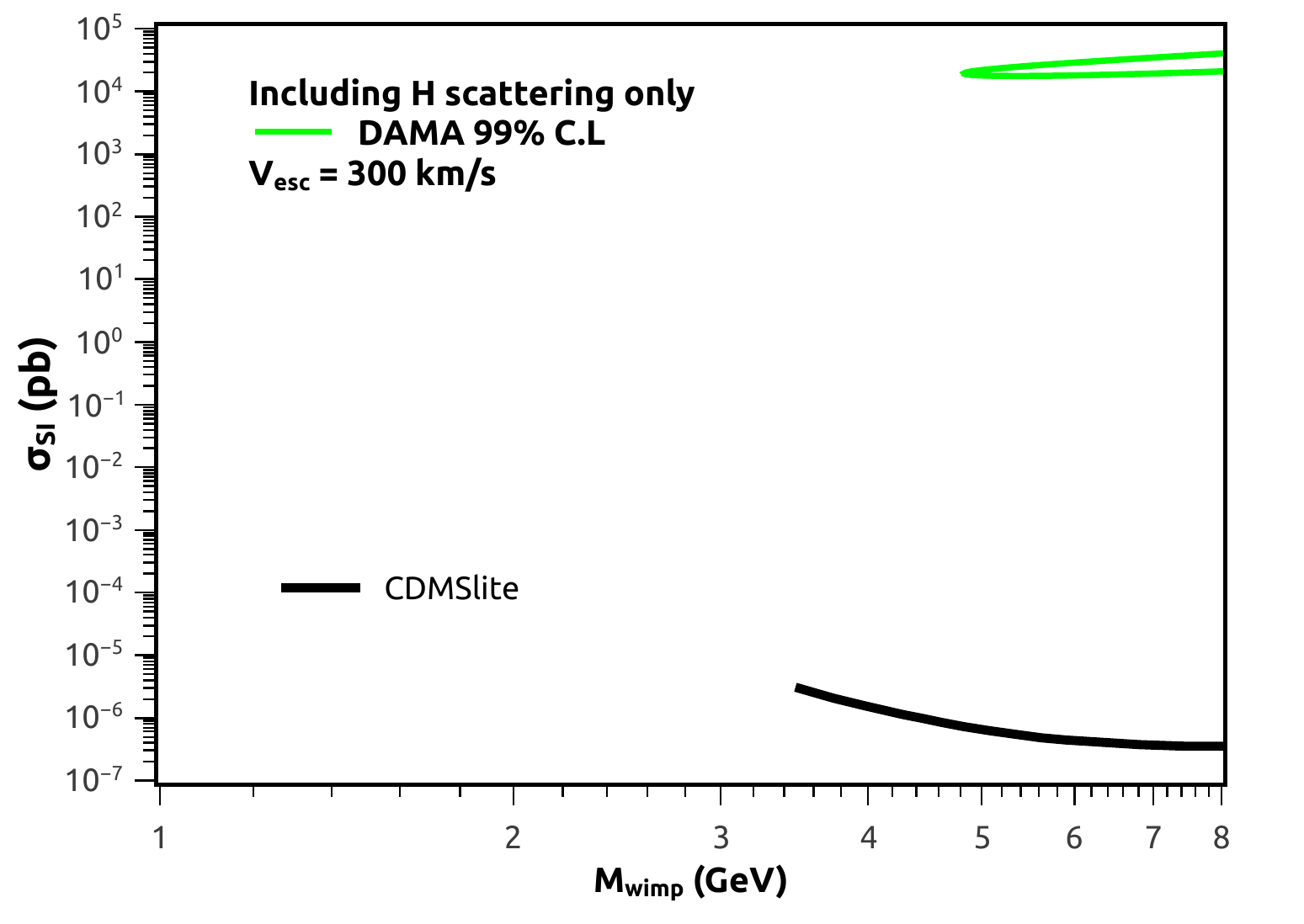}\quad\includegraphics[width=0.75\columnwidth]{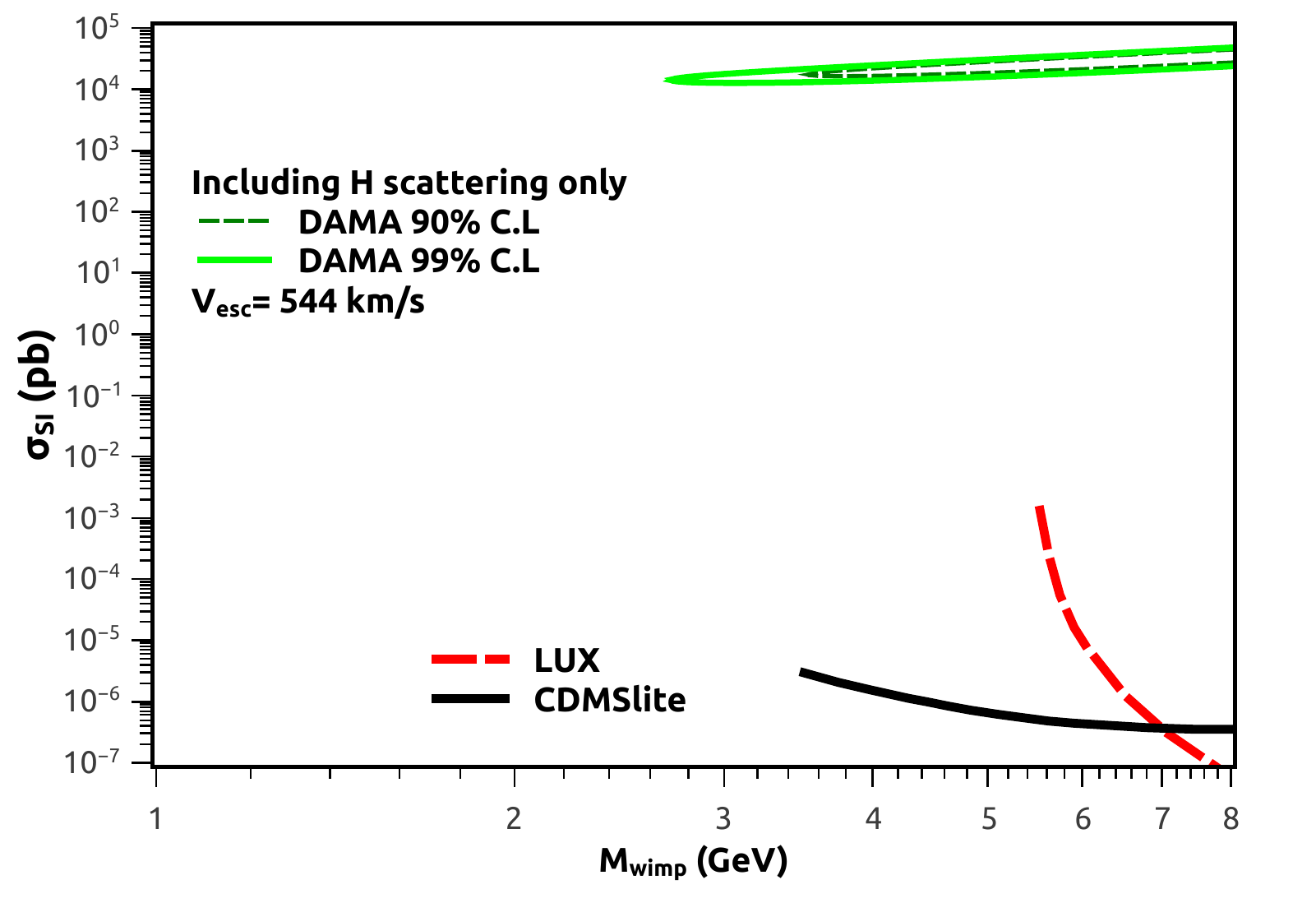}\quad\includegraphics[width=0.73\columnwidth]{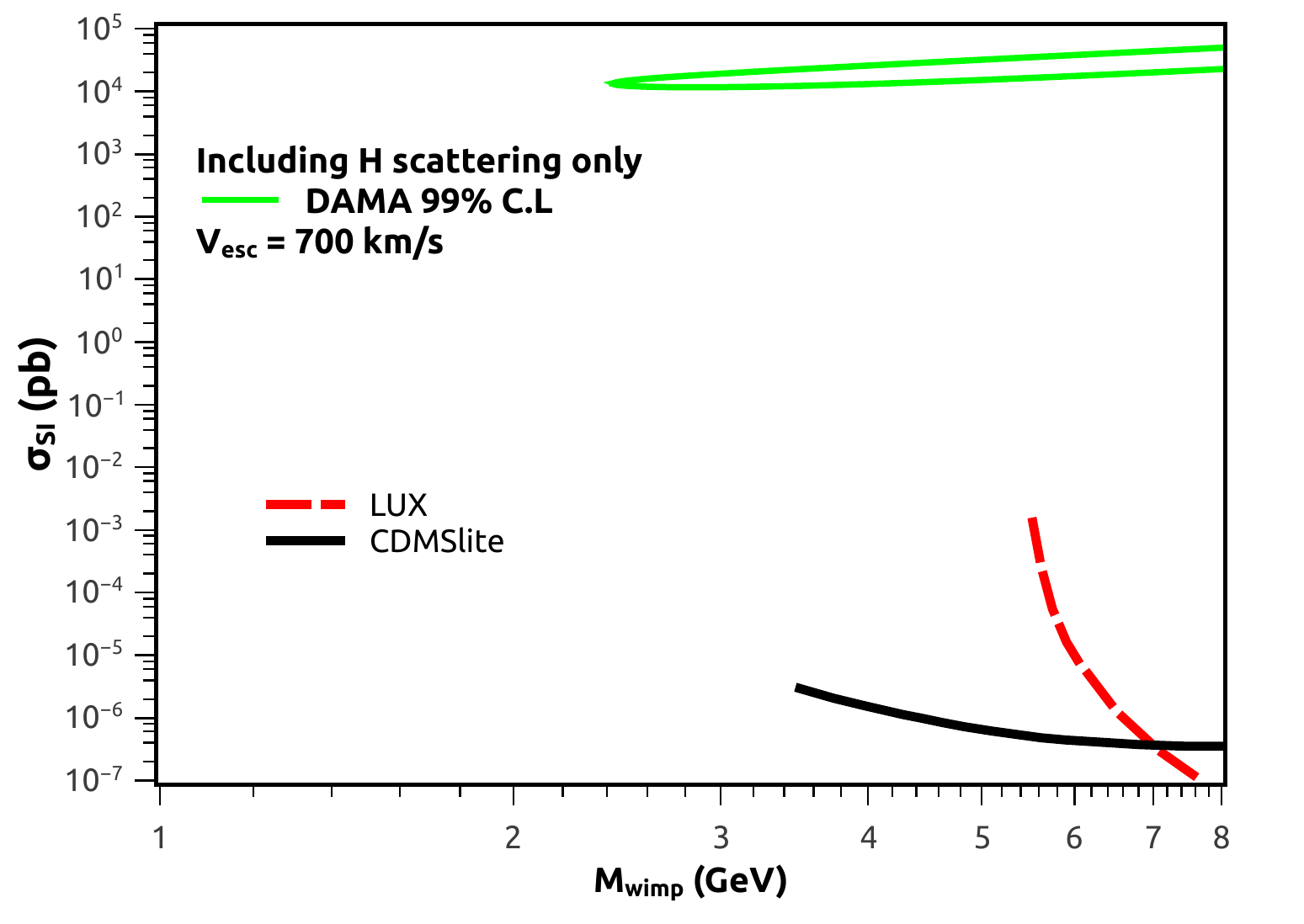}}
\caption{The best fit regions under the assumption of exclusive WIMP-Hydrogen scattering (no Na scattering) using an escape velocity, from left to right, of 300, 544 and 700 km/s.We use a $1$~ppm contamination of Hydrogen in NaI(Tl) crystals. The $90\%$~C.L region is excluded for all cases by many orders of magnitude by CDMSlite; a small mass window $2.5-3.5$~GeV is still consistent with current limits for large enough escape velocities, but is inconsistent with other bounds (see the text).}
\label{Graph1}
\end{figure*}

\section{Results}

Ref.~\cite{claim} assumes that dark matter particles might only scatter off of Hydrogen, but not off of Na. This, if at all possible, would definitely make for highly peculiar particle properties. We examine this possibility and compare the best fit regions with available constraints, in fig.~~\ref{Graph1}. There, we assume a quenching factor of one and a standard Mawellian velocity distribution, with three different escape velocities: $v_{\rm esc}=300$, 544 and 700 km/s. The reason for testing different escape velocities is to check the effect of changing the standard assumptions regarding the standard halo model.
Notice that in the leftmost panel in fig.~~\ref{Graph1}, only CDMSlite bound applies due to the very low escape velocity.

Fig.~\ref{Graph1} illustrates that even if one neglected Na recoils and considered Hydrogen scattering only, the DAMA signal could be fit satisfactorily only down to a mass of 2.5 GeV, at the 99\% C.L.. The $90\%$~C.L region is excluded for all cases by many orders of magnitude by CDMSlite;  the small mass window $2.5-3.5$~GeV is still consistent with current limits only for large enough escape velocity. Notice that smaller quenching factors (which are much more likely) would shift the favored region to even higher masses, and strengthen the bounds coming from other experiments. Higher masses are vastly excluded by orders of magnitude, as also shown in fig.~\ref{Graph1}. Even only at a mass of 3 GeV the large cross sections required would be in tension with the CDMSlite results (assuming that the WIMP would scatter off of Ge). In the very small mass region (2.5 to 3.5 GeV) such large cross sections are inconsistent with high-altitude detector results and with heat flow measurements, as we discuss below. %, an experiment which features an energy threshold of $0.84$~keVnr \cite{cdmslite}.

%%%% SP EDITED TILL HERE

\begin{figure*}[t]
\centering
\mbox{\hspace*{-1cm}\includegraphics[width=0.73\columnwidth]{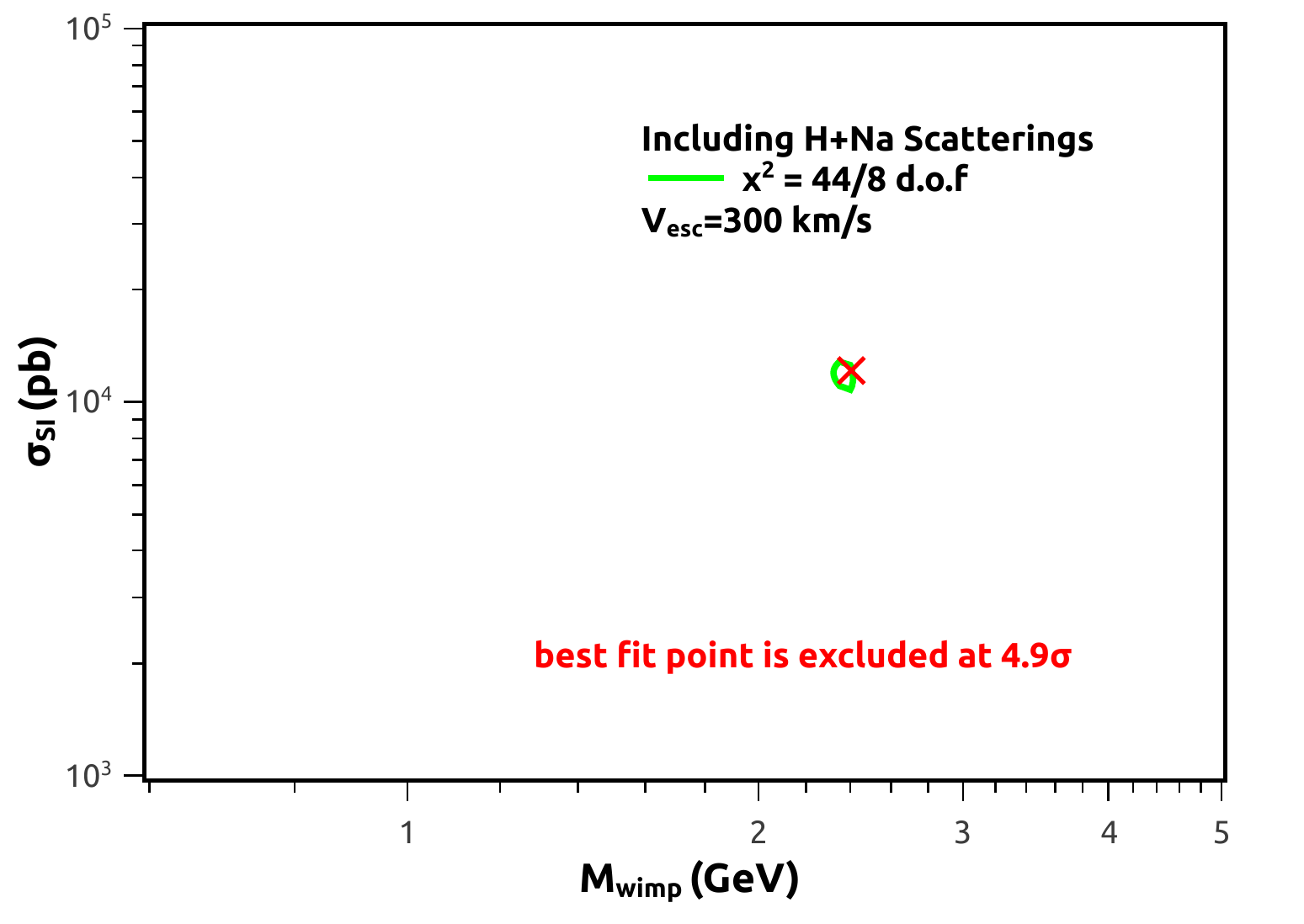}\quad\includegraphics[width=0.73\columnwidth]{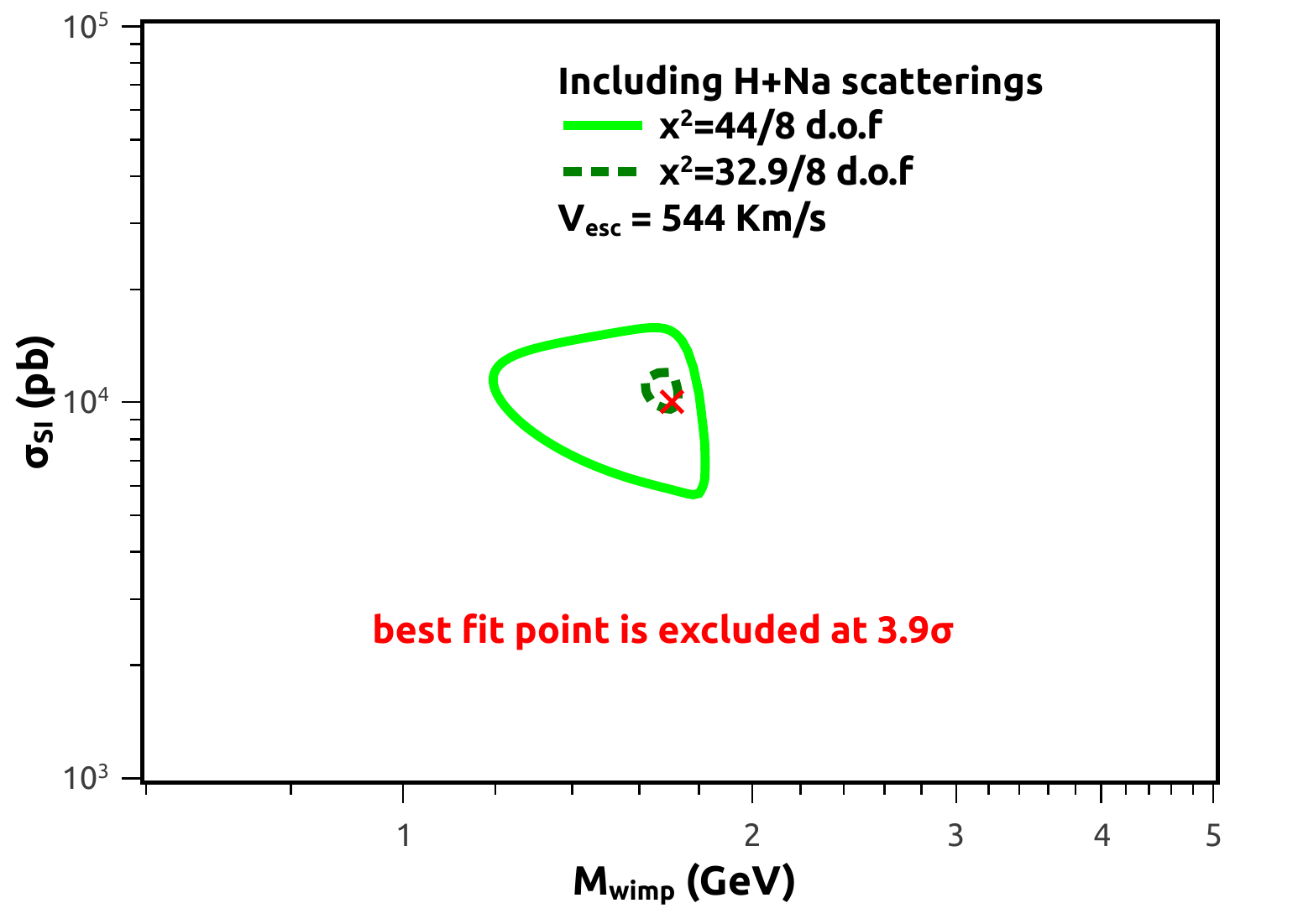}\quad\includegraphics[width=0.73\columnwidth]{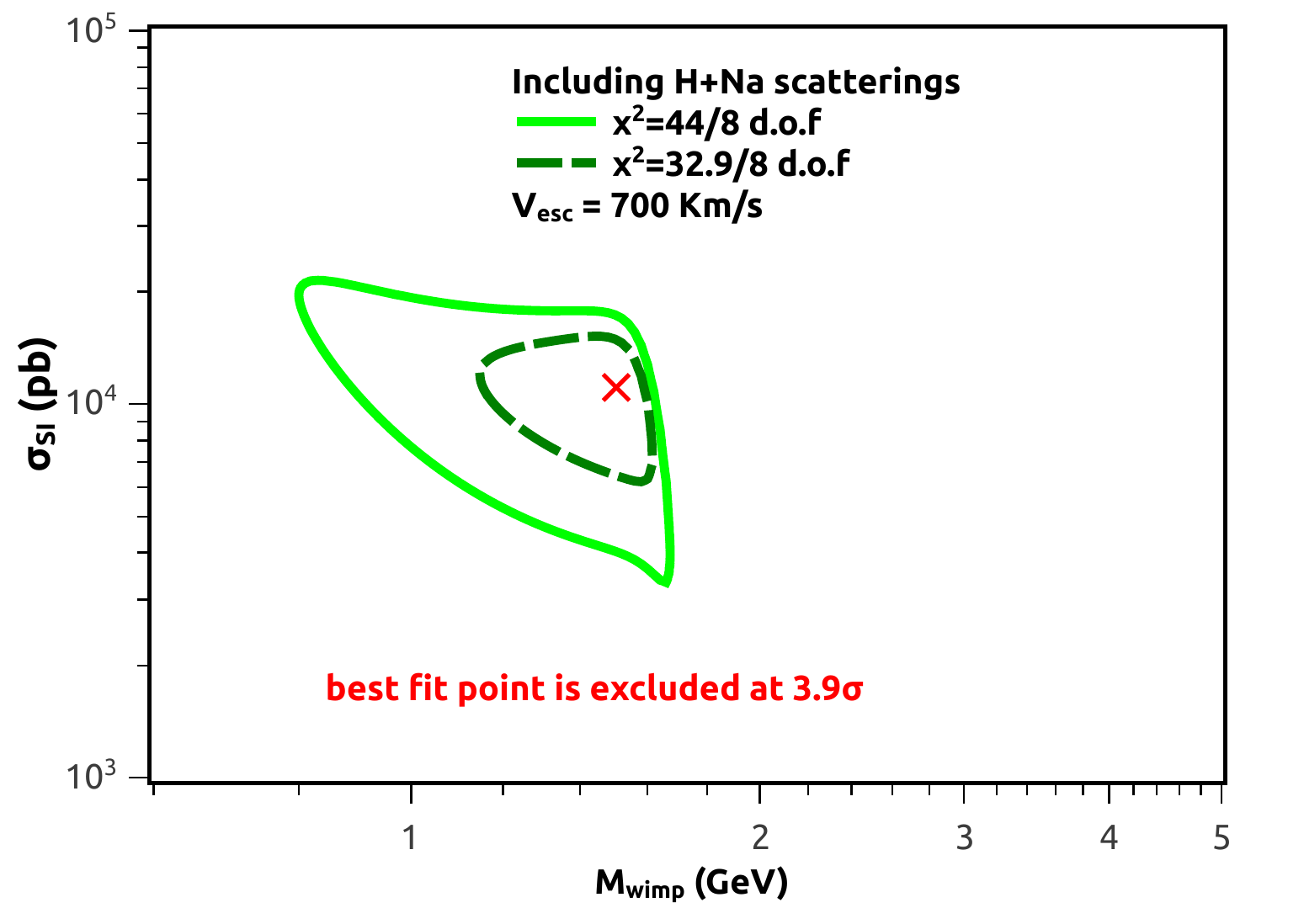}}
\caption{The best fit regions including both H and Na scattering using $V_{esc}=300$, 544 and 700~km/s (from left to right). The favored mass region is not excluded by other direct detection experiments, but the best-fit input parameters provide a very poor fit to the DAMA/LIBRA data. For example, even the best-fit point (indicated with a red cross $\sigma =10^4$~pb and $M=1.7$~GeV) is excluded at the $3.9\sigma$ level for $V_{\rm esc}=544$ and 700 km/s (the fit for $V_{\rm esc}=544$ km/s is even worse). All contours are goodness of fit contours using the binning scheme described in Ref. \cite{chrisbin}.}
\label{Graph2}
\end{figure*}

We now consider the more realistic case where scattering happens off of nuclei in both H and Na. In fig.~\ref{Graph2} we show best-fit contours of constant $\chi^2$ values (44 and 32.9), as well as the best fit point (red X). The favored mass region is not excluded by other direct detection experiments, but the best-fit input parameters provide a very poor fit to the DAMA/LIBRA data. For example, even the best-fit point (indicated with a red cross $\sigma =10^4$~pb and $M=1.7$~GeV) is excluded at the $3.9\sigma$ level for $V_{\rm esc}=544$ and 700 km/s.  Even if one used extremely low values for the WIMP escape velocity in an attempt to suppress Na scatterings (left panel, $V_{\rm esc}=300$ km/s), the predicted signal provides a very poor fit to the modulation observed by DAMA/LIBRA.% and for this reason only a 99\%~CL region was drawn in fig~\ref{Graph1_300}. Non-standard halo models can predict higher escape velocities, larger than 544km/s, and hence such a setup would worsen the very light scenario shown in Fig.\ref{Graph1} because there would be even more higher speed particles.
%\begin{figure}[t]
%\centering
%\includegraphics[scale=0.6]{result1.pdf}
%\includegraphics[width=1\columnwidth]{result1new_vesc300.pdf}
%\caption{The 99\% best fit region under the assumption of exclusive WIMP-Hydrogen scattering (no Na scattering) using a extremely low escape velocity of $300$~Km/s. Higher scape velocities will worsen the scenario, because more WIMP-Na events scatterings will happen. The fit is very poor at this point and for this reason just a 99\%C.L region has been shown. As in Ref.~\cite{claim}, we used a $1$~ppm contamination of Hydrogen in NaI(Tl) crystals. The $90\%$~C.L region is excluded by many orders of magnitude by CDMSlite and just a small mass window $2.5-3.5$~GeV is still consistent with current limits.}
%\label{Graph1_300}
%\end{figure}
%
%Is this modulation signal due to dark matter scatterings? Seemingly not. But, it is not due to a $1$~GeV canonical WIMP. 
%\clearpage

%
%
%If WIMP-Na scattering is included, i.e if, as expected, the WIMP scattered both from Na and H, one would find the result shown in fig.~\ref{Graph2}-\ref{Graph2_300} for $v_{esc}=544$~km/s and $v_{esc}=300$~km/s respectively. In fig.~\ref{Graph2} we show best-fit regions with $\chi^2=32.9/8$~d.o.f and $\chi^2=44/8$~d.o.f. It is clear from the $\chi^2$ contours that this setup provides a very poor fit to the DAMA modulation. Indeed, the best fit point is excluded at the $3.9\sigma$ level. All other points are excluded at even higher confidence level. 

We thus conclude that a very light WIMP as a candidate to explain the DAMA modulation is excluded at almost $\sim 4 \sigma$.  In this configuration, in fact, the WIMPs in the tail of the velocity distribution are responsible for scattering off of Na, and produce an extremely large number of events, therefore entailing a poor fit to the modulation signal. We verified that even assuming a very small escape velocity does not improve the situation: while the best fit region is shifted to lower masses, the fit to the modulation signal is overall worse than with higher escape velocities.

It is crucial to note that there exist other lines of argument why a 1 GeV WIMP with a large scattering cross section is strongly disfavored as an explanation to the DAMA/LIBRA modulation signal: high-altitude detectors and Earth heating. A light and strongly interacting particle would violate bounds from high-altitude detectors \cite{Rich:1987st} which unlike underground detectors are not shielded. With such large cross sections, large portions of the favored parameter space in the light WIMP scenario is in disagreement with observation.

\begin{figure}[]
\centering
\includegraphics[width=0.8\columnwidth]{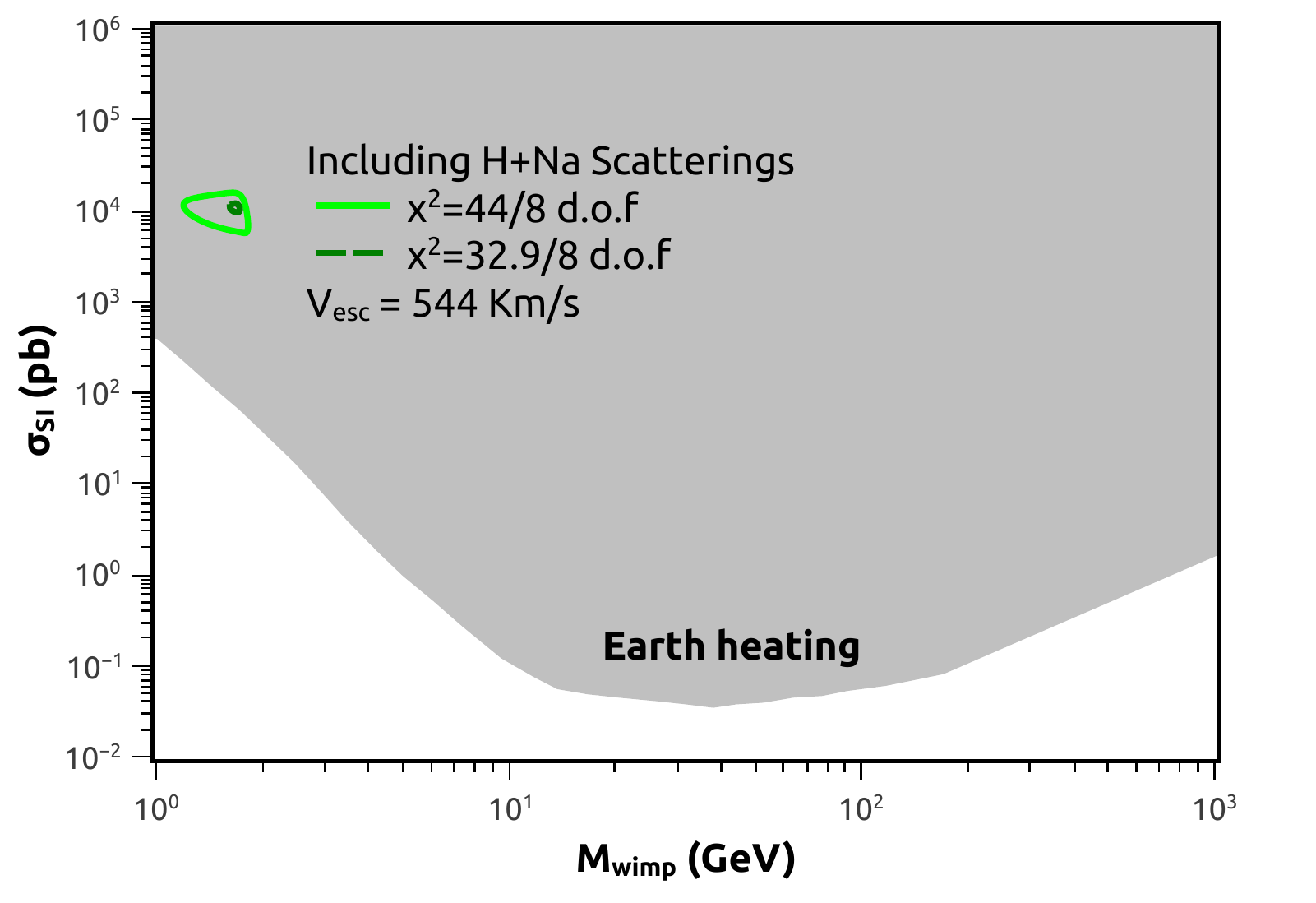}
\caption{Exclusion region based on the capture of Dark Matter particles in the Earth and subsequent annihilation, producing an excess over the measured heat flow of Earth \cite{Mack:2007xj}. We show with green contours the best-fit DAMA region for an escape velocity of 544 km/s.}
\label{Graph3}
\end{figure}

Even stronger bounds come from precision measurements of the Earth heat flow. When the dark matter scattering rate is sufficiently large, the energy deposition by dark matter self-annihilation products captured by the Earth gravitational potential would grossly exceed the measured heat of Earth. A largely model-independent constraint on the dark matter scattering cross section with nucleons can be placed \cite{Mack:2007xj}, under the assumption that annihilation occurs. We display for reference the resulting constraints from Earth heating as the shaded region in Fig.\ref{Graph3}: the excluded region completely rules out the best-fit DAMA region even when only Hydrogen scattering are taken into account (we use here a quenching factor of one and escape velocity of $544$~km/s, but our results are largely independent of these two assumptions).

\section{Conclusions}

In this analysis we demonstrated that a light, $\mathcal{O}$(1) GeV WIMP scattering off of Hydrogen nuclei from residual OH contamination in the DAMA NaI(Tl) crystals is a highly disfavored scenario for the following reasons: 

(i) under the unmotivated assumption that the WIMP only scatter off of Hydrogen but not Na, we showed that the favored mass range would place the proton-WIMP cross section in a region grossly ruled out by current experiments; 

(ii) if scattering off of Na exists concurrently with H, the resulting signal does not provide a good fit to the DAMA signal, and is ruled out to almost the 4$\sigma$ level;

(ii) if scattering off of Ge exists concurrently with H, then CDMSlite limits eliminate the favored WIMP candidate mass;

(iv) indirect constraints such as limits from high-altitude detectors and from the Earth heat flow are generically incompatible with the proposed mass and cross section.\\

\section*{Acknowledgments}
The authors thank Ritoban Basu, Will Shepherd, Patrick Draper for relevant discussions and Chris Kelso for reading the manuscript and discussing the results. SP is partly supported by Department of Energy Award SC0010107, and FSQ by Department of Energy Award SC0010107 and National Council for Scientific and Technological Development (CNPq).


\begin{thebibliography}{99}\frenchspacing

\bibitem{claim} 
  J.~Va'vra,
   ``{\em A New Possible Way to Explain DAMA Results},''
  arXiv:1401.0698 [astro-ph.GA].
  %%CITATION = ARXIV:1401.0698;%%

\bibitem{Agnese:2013rvf} 
  R.~Agnese {\it et al.}  [CDMS Collaboration],
  %``Dark Matter Search Results Using the Silicon Detectors of CDMS II,''
  %Submitted to: Phys.Rev.Lett.
  [arXiv:1304.4279 [hep-ex]].
  %%CITATION = ARXIV:1304.4279;%%

\bibitem{Aalseth:2010vx}
  C.~E.~Aalseth {\it et al.} [CoGeNT Collaboration],
  %``Results from a Search for Light-Mass Dark Matter with a P-type Point 
  Phys.\ Rev.\ Lett.\  {\bf 106}, 131301 (2011).
  [arXiv:1002.4703 [astro-ph.CO]].


\bibitem{Aalseth:2011wp}
  C.~E.~Aalseth, P.~S.~Barbeau, J.~Colaresi, J.~I.~Collar, J.~Diaz Leon, J.~E.~Fast, N.~Fields, T.~W.~Hossbach {\it et al.},
  %``Search for an Annual Modulation in a P-type Point Contact Germanium
  Phys.\ Rev.\ Lett.\  {\bf 107}, 141301 (2011).
  [arXiv:1106.0650 [astro-ph.CO]].  
  
%\cite{Angloher:2011uu}
\bibitem{cogentnew}
C.~E.~Aalseth {\it et al.},
  %``Search for An Annual Modulation in Three Years of CoGeNT Dark Matter Detector Data,''
  arXiv:1401.3295 [astro-ph.CO].
  %%CITATION = ARXIV:1401.3295;%%

\bibitem{Angloher:2011uu}
  G.~Angloher, M.~Bauer, I.~Bavykina, A.~Bento, C.~Bucci, C.~Ciemniak, G.~Deuter, F.~von Feilitzsch {\it et al.},
  %``Results from 730 kg days of the CRESST-II Dark Matter Search,''
  [arXiv:1109.0702 [astro-ph.CO]].
  
  
\bibitem{Bernabei:2013qx} 
  R.~Bernabei, P.~Belli, A.~Di Marco, F.~Cappella, A.~d'Angelo, A.~Incicchitti, V.~Caracciolo and R.~Cerulli {\it et al.},
  %``DAMA/LIBRA results and perspectives,''
  arXiv:1301.6243 [astro-ph.GA].
  
\bibitem{weiner}
P.~J.~Fox, G.~Jung, P.~Sorensen and N.~Weiner,
  %``Dark Matter in Light of LUX,''
  arXiv:1401.0216 [hep-ph].
  %%CITATION = ARXIV:1401.0216;%%
  %1 citations counted in INSPIRE as of 16 Jan 2014

\bibitem{xenon}
E.~Aprile {\it et al.}  [XENON100 Collaboration],
  %``Dark Matter Results from 225 Live Days of XENON100 Data,''
  Phys.\ Rev.\ Lett.\  {\bf 109}, 181301 (2012)
  [arXiv:1207.5988 [astro-ph.CO]].
  %%CITATION = ARXIV:1207.5988;%%

\bibitem{LUX} LUX Collaboration, [arXiv:1310.8214].  
    
\bibitem{isospin}
J.~L.~Feng, J.~Kumar, D.~Marfatia and D.~Sanford,
  %``Isospin-Violating Dark Matter,''
  Phys.\ Lett.\ B {\bf 703}, 124 (2011)
  [arXiv:1102.4331 [hep-ph]].

\bibitem{cdmslite} CDMSlite Collaboration,[arXiv:1309.3259].

\bibitem{supercdms} SUPERCDMS Collaboration,[arXiv:1402.7137].

\bibitem{chrisbin} Chris Kelso, Pearl Sandick, Christopher Savage, JCAP 1309 (2013) 022, 
[arxiv:1306.1858].

\bibitem{modulationreview} Katherine Freese, Mariangela Lisanti, Christopher Savage, Rev.Mod.Phys. 85 (2013) 1561-1581, [arXiv:1209.3339].

\bibitem{streamsDAMA} C. Savage, K. Freese and P. Gondolo, Phys. Rev. D 74, 043531 (2006) [astro-ph/0607121].

\bibitem{Rich:1987st} 
  J.~Rich, R.~Rocchia and M.~Spiro,
  %``A Search for Strongly Interacting Dark Matter,''
  Phys.\ Lett.\ B {\bf 194}, 173 (1987).
  %%CITATION = PHLTA,B194,173;%%
  %31 citations counted in INSPIRE as of 16 Jan 2014

\bibitem{Mack:2007xj} 
  G.~D.~Mack, J.~F.~Beacom and G.~Bertone,
  %``Towards Closing the Window on Strongly Interacting Dark Matter: Far-Reaching Constraints from Earth's Heat Flow,''
  Phys.\ Rev.\ D {\bf 76}, 043523 (2007)
  [arXiv:0705.4298 [astro-ph]].
  %%CITATION = ARXIV:0705.4298;%%
  %54 citations counted in INSPIRE as of 16 Jan 2014


\end{thebibliography}
\end{document}